\documentclass[twocolumn,pra, aps, superscriptaddress,longbibliography,showpacs,amsmath,amssymb,floatfix]{revtex4-2}       
\usepackage{graphics}
\usepackage{amssymb}
\usepackage{ulem}
\usepackage[section]{placeins}
\usepackage{amsmath}
\usepackage{epsfig,braket,mathrsfs,multirow}
\usepackage{color,soul}
\usepackage{tabu}
\usepackage{mathtools}
\usepackage[colorlinks,linkcolor=blue,anchorcolor=blue,citecolor=blue,urlcolor=blue]{hyperref}
\usepackage{physics}
\usepackage{float}
\usepackage{diagbox}
\usepackage{subfiles}

\begin{document}


\title{Critical fluctuation and noise spectra in two-dimensional Fe$_{3}$GeTe$_{2}$ magnets}

\author{Yuxin Li}
\thanks{These authors contributed equally to this work.}
\affiliation{CAS Key Laboratory of Microscale Magnetic Resonance and School of Physical Sciences, University of Science and Technology of China, Hefei 230026, China}
\affiliation{Anhui Province Key Laboratory of Scientific Instrument Development and Application, University of Science and Technology of China, Hefei 230026, China}
\affiliation{Hefei National Laboratory, University of Science and Technology of China, Hefei 230088, China}

\author{Zhe Ding}
\thanks{These authors contributed equally to this work.}
\affiliation{CAS Key Laboratory of Microscale Magnetic Resonance and School of Physical Sciences, University of Science and Technology of China, Hefei 230026, China}
\affiliation{Anhui Province Key Laboratory of Scientific Instrument Development and Application, University of Science and Technology of China, Hefei 230026, China}

\author{Chen Wang}
\thanks{These authors contributed equally to this work.}
\affiliation{Hefei National Laboratory, University of Science and Technology of China, Hefei 230088, China} 
\affiliation{International Center for Quantum Design of Functional Materials (ICQD), Hefei National Research Center for Physical Science at the Microscale, University of Science and Technology of China, Hefei 230026, China}

\author{Haoyu Sun}
\affiliation{CAS Key Laboratory of Microscale Magnetic Resonance and School of Physical Sciences, University of Science and Technology of China, Hefei 230026, China}
\affiliation{Anhui Province Key Laboratory of Scientific Instrument Development and Application, University of Science and Technology of China, Hefei 230026, China}

\author{Zhousheng Chen}
\affiliation{CAS Key Laboratory of Microscale Magnetic Resonance and School of Physical Sciences, University of Science and Technology of China, Hefei 230026, China}
\affiliation{Anhui Province Key Laboratory of Scientific Instrument Development and Application, University of Science and Technology of China, Hefei 230026, China}

\author{Pengfei Wang}
\affiliation{CAS Key Laboratory of Microscale Magnetic Resonance and School of Physical Sciences, University of Science and Technology of China, Hefei 230026, China}
\affiliation{Anhui Province Key Laboratory of Scientific Instrument Development and Application, University of Science and Technology of China, Hefei 230026, China}
\affiliation{Hefei National Laboratory, University of Science and Technology of China, Hefei 230088, China} 

\author{Ya Wang}
\affiliation{CAS Key Laboratory of Microscale Magnetic Resonance and School of Physical Sciences, University of Science and Technology of China, Hefei 230026, China}
\affiliation{Anhui Province Key Laboratory of Scientific Instrument Development and Application, University of Science and Technology of China, Hefei 230026, China}
\affiliation{Hefei National Laboratory, University of Science and Technology of China, Hefei 230088, China} 

\author{Ming Gong}
\email{gongm@ustc.edu.cn}
\affiliation{Hefei National Laboratory, University of Science and Technology of China, Hefei 230088, China} 
\affiliation{CAS Key Laboratory of Quantum Information,University of Science and Technology of China, Hefei 230026, China}

\author{Hualing Zeng}
\email{hlzeng@ustc.edu.cn}
\affiliation{Hefei National Laboratory, University of Science and Technology of China, Hefei 230088, China} 
\affiliation{International Center for Quantum Design of Functional Materials (ICQD), Hefei National Research Center for Physical Science at the Microscale, University of Science and Technology of China, Hefei 230026, China}

\author{Fazhan Shi}
\email{fzshi@ustc.edu.cn}
\affiliation{CAS Key Laboratory of Microscale Magnetic Resonance and School of Physical Sciences, University of Science and Technology of China, Hefei 230026, China}
\address{Anhui Province Key Laboratory of Scientific Instrument Development and Application, University of Science and Technology of China, Hefei 230026, China}
\affiliation{Hefei National Laboratory, University of Science and Technology of China, Hefei 230088, China}
\affiliation{School of Biomedical Engineering and Suzhou Institute for Advanced Research, University of Science and Technology of China, Suzhou 215123, China}

\author{Jiangfeng Du}
\affiliation{CAS Key Laboratory of Microscale Magnetic Resonance and School of Physical Sciences, University of Science and Technology of China, Hefei 230026, China}
\affiliation{Anhui Province Key Laboratory of Scientific Instrument Development and Application, University of Science and Technology of China, Hefei 230026, China}
\affiliation{Hefei National Laboratory, University of Science and Technology of China, Hefei 230088, China}
\affiliation{Institute of Quantum Sensing and School of Physics, Zhejiang University, Hangzhou 310027, China}

\begin{abstract}
Critical fluctuations play a fundamental role in determining the spin orders for low-dimensional quantum materials, especially for recently discovered two-dimensional (2D) magnets. Here we employ the quantum decoherence imaging technique utilizing nitrogen-vacancy centers in diamond to explore the critical magnetic fluctuations and the associated temporal spin noise in van der Waals magnet $\rm{Fe_{3}GeTe_{2}}$. We show that the critical fluctuation contributes to a random magnetic field  characterized by the noise spectra, which can be changed dramatically near the critical temperature $T_c$. A theoretical model to describe this phenomenon is developed, showing that the spectral density is characterized by a $1/f$ noise near the $T_c$, while away from this point it behaves like a white noise. The crossover at a certain temperature between these two situations is determined by changing of the distance between the sample and the diamond. This work provides a new way to study critical fluctuation and to extract some of the critical exponents, which may greatly deepen our understanding of criticality in a wide range of physical systems.
\end{abstract} 
\maketitle

Phase transitions and the associated fluctuations near the critical points are two major themes in condensed matter physics \cite{Elliott1983Magneticbook,Hohenberg1977theory,sachdev2011Quantum}. It has been known since the 1960s  
\cite{Kadanoff1967Static,Fisher1974RG,WILSON197475} that the phase transition is accompanied by fluctuation near the critical point from the long-range correlations, yielding physics quantitatively different from the mean-field theory. As a result, singularities in magnetization, susceptibility, and specific heat following some power laws can be observed, and their critical exponents have been extracted \cite{Kadanoff1967Static,Liu2017CriticalFGT1,Tan2018FGT,Fei20182DFGT}, which are standard tools to characterize phase transitions. For magnets, although these quantities have been well understood, the spatial and temporal fluctuations \cite{SalathIsing2015,Binder1997MC}, which can be regarded as some kind of noise are not directly identified in experiments. 

\begin{figure}[ht]
\centering
\includegraphics[width=1\linewidth]{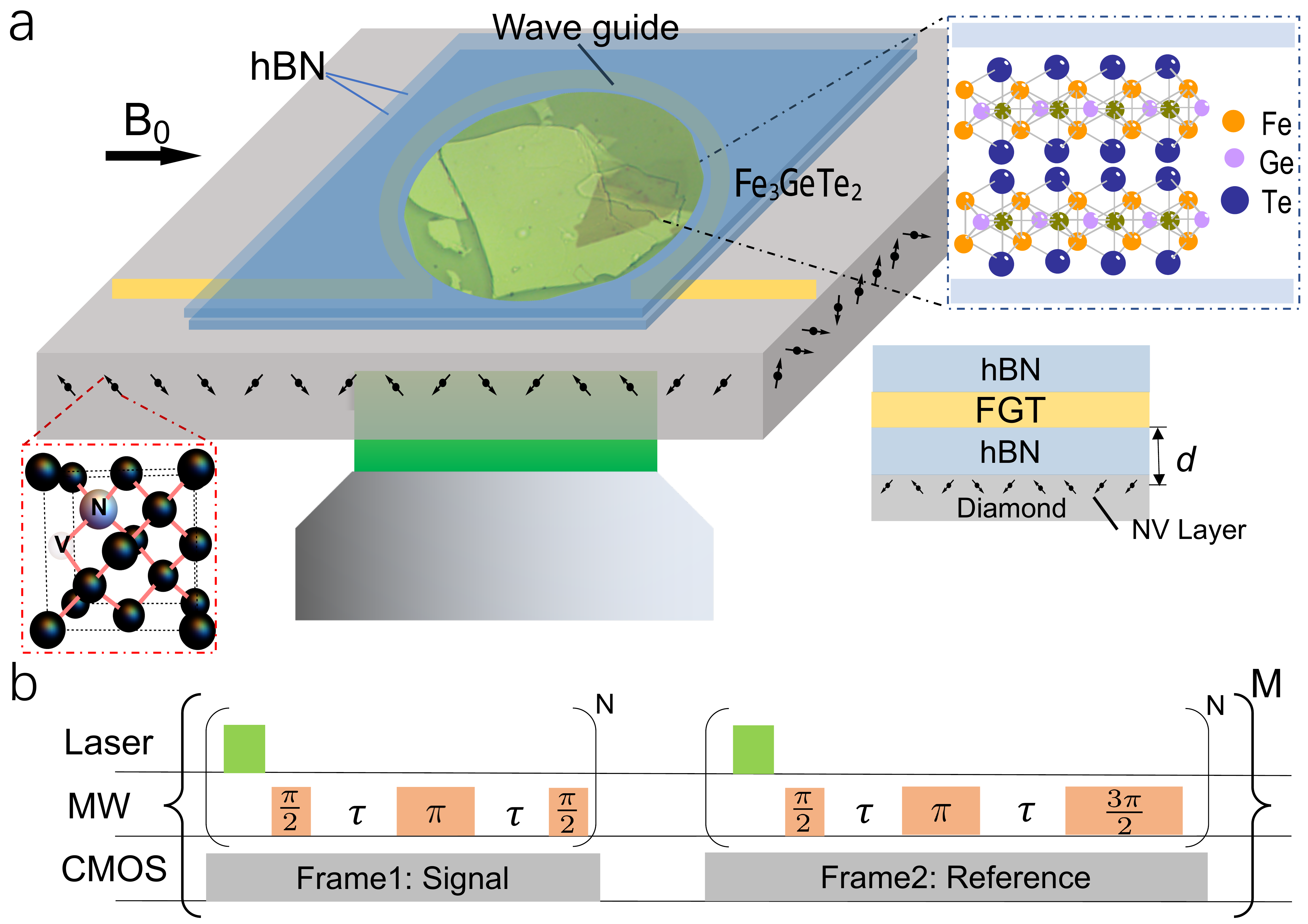}
\caption{{\bf Experiment setup and pulse sequences}. (a) The exfoliated FGT flakes are transferred to a [100] oriented diamond chip with  single-layer NV centers. Microwave excitation is radiated through an $\Omega$-shaped antenna, and the magnetic field $B_{0}$ is applied in-plane. hBN substrate layers are transferred above and below the FGT flakes. The inset illustrates the crystal structure of FGT and NV centers, with $d$ being the distance between the FGT flakes and the NV centers. (b) Hahn-echo pulse sequences for coherence detection on wide-field microscopy. The laser pulses initialize the spin state to $\ket{0}$ and readout the final state. The first microwave pulse drives the spin of NV centers to the superposition state, then the $\pi$ pulse decouples the noise whose frequency mismatches the pulse interval $\tau$. The final pulse is configured to be either $\frac{\pi}{2}$ or $\frac{3\pi}{2}$ to transfer the coherence to the population of the state $\ket{0}$ or $\ket{1}$. The fluorescence of these two populations is recorded by CMOS as the signal frame and the reference frame, respectively. In the experiment, each $\tau$ is averaged by $M \sim 100 - 200$ measurements.}
\label{fig-fig1}
\end{figure}

Imagine an experiment with a sensor placed above a magnet, which can instantaneously measure the temporal perturbation caused by the fluctuation during phase transition, then we can ask two fundamental questions. Firstly, what kind of physics can be reflected from these observations; and Secondly, what is the relation between the observation and the critical physics in the sample? Take the 
ferromagnetic to paramagnetic transition as an example, near the critical temperature, the spin fluctuation can be regarded as some kind of noise with some distributions in the frequency domain. The noise measured by the sensor can be fully characterized by some power spectra density $S_{B}(\omega)$, which is closely related to the spectra density of magnetization $S_{M}(\omega)$ in the sample. From Ref. \cite{Chen2007Measurement} the noise spectra depends strongly on $(T-T_c)/T_{c}$, where $T$ and $T_c$ are the temperature and critical temperature, respectively. Therefore, the above two questions are fundamentally related to the measurement of the noise spectra, which will deepen our understanding of criticality \cite{Kwan1993exponent,Chen2007Measurement,Chen2009telenoise}. 

This work uses the nitrogen-vacancy (NV) center in diamond as a quantum sensor \cite{degen2014NV,jiangfeng2024rmp} to measure the critical spin fluctuation in two-dimensional (2D) magnets \cite{Tan2018FGT,Kurebayashi2022magnetismvdW}. By applying dynamical decoupling pulse sequences to  the  NV centers, we can extract the noise spectra generated by the critical fluctuation from the quantum decoherence process. Since the coherence time is several microseconds, only the noise frequency at MHz contributes to this observation. Two major conclusions are reached. Firstly, we find that the decoherence rate is greatly enhanced near the critical temperature due to the critical fluctuation. We contribute this result to the change of spectra density near the critical point.  Secondly, we develop a theoretical model to describe the change in decoherence rate, which depends on the 
noise spectra and the separation between the sensor and sample. Furthermore, the crossover condition from $1/f$ noise to white noise is given in the experiment explicitly. Our results pave the way to understanding the critical fluctuation in terms of noise spectra in the other materials. 

We study the critical fluctuation of $\rm{Fe_{3}GeTe_{2}}$, hereafter termed as FGT, which exhibits magnetic phase transition. This is an exfoliable vdW magnet exhibiting robust 2D ferromagnetism with strong perpendicular anisotropy even when thinned down to monolayer  \cite{Deiseroth2006Fe3GeTe2AN,Tan2018FGT,Fei20182DFGT}. For the bulk material, the $T_{c}$ has been determined to be about 210 K \cite{Liu2017CriticalFGT1,Deng2018GateRTFGTb}. When thinned down to monolayer, with the increasing fluctuation effect, the $T_c$ is dramatically reduced \cite{Zhang2001ThicknessTc}. The critical fluctuation near the critical point is the major concern of this work. To this end, we set up a cryogenic wide-field microscope system based on 
 ensemble NV centers \cite{scholten2021widefield,zschennv} to carry out all experiments. The experiment configuration is illustrated in Fig. \ref{fig-fig1} (a). A linear polarized 532 nm laser is employed for initialization and readout of the state of NV centers and microwave is radiated by the antenna to manipulate the quantum state of NV centers. An in-plane magnetic field is applied to split the degeneracy of the $\ket{\pm 1}$ states of the NV centers and to avoid perturbing the anisotropy of FGT. Hexagonal boron nitride (hBN) substrates are transferred above and below the FGT flakes to prevent oxidation and to adjust the distance $d$ between the FGT and NV centers, respectively. More details about the setup are presented in Section \uppercase\expandafter{\romannumeral 1} of  Supplementary Material \cite{Supp_Mat}.

The NV centers, serving as quantum sensor, have found extensive applications in various fields  \cite{Casola2018Probing,Gross2017BFO,Song2021nvtwistCrI3,Bhattacharyya2024MeissnerSC,Masaya2024SWNV}. Through the optically detected magnetic resonance technique, the static magnetic field  can be measured by continuous wave (CW) spectrum \cite{Balasubramanian2008}. Meanwhile, based on the quantum coherence and relaxation properties of NV centers, dynamic decoupling techniques have been developed \cite{Suter2016Colloquium,G.deLange2010Universal,Du2009,Wang2012ComparisonNV} to collect signals in the frequency domain for sensing dynamic signals \cite{Fazhanshi2015protein,Wang2023Visualization}. Indeed, relevant critical fluctuation phenomena have been detected in vdW magnetic materials, such as $\rm{Fe_{3}GeTe_{2}}$ \cite{Huang2022WideField}, $\rm{MnBi_{2}Te_{4}(Bi_{2}Te_{3})_{n}}$ \cite{McLaughlin2022QuantumImaging}, CrPS$_{4}$ \cite{Huang2023CPS4} and twisted double trilayer $\rm{CrI_{3}}$ \cite{huang2023revealing}, mainly utilizing the relaxation properties of the NV centers. The relaxation time $T_1$ of NV centers is primarily influenced by noise with frequency resonant with the energy levels, typically around 2.87 GHz. The increasing of relaxation rate reflects the increasing of spin fluctuation at the critical temperature. Furthermore, sub-Hertz flips of the magnetic domain have been observed in our previous work near $T_c$ \cite{Jin2020Imaging, Wang2023Thermal}. Thus the NV centers can be used to measure the fluctuations at different time scales. 

In contrast to the previous experiments, our work utilizes the quantum coherence of NV centers to investigate the critical spin fluctuation. The details for the NV centers and FGT samples are presented in Section \uppercase\expandafter{\romannumeral 2} and \uppercase\expandafter{\romannumeral 3} of  Supplementary Material \cite{Supp_Mat}. The Hahn-echo pulse sequence is shown in Fig. \ref{fig-fig1} (b). The weight function of this pulse sequence is narrow banded \cite{Degen2017QuantumSensing}, thus only noise with frequency matches the interval of the pulse sequence contributes to the decoherence, leading to the change of $T_2$ in the sensor. 

\begin{figure}[ht]
		\centering
		\includegraphics[width=1\linewidth]{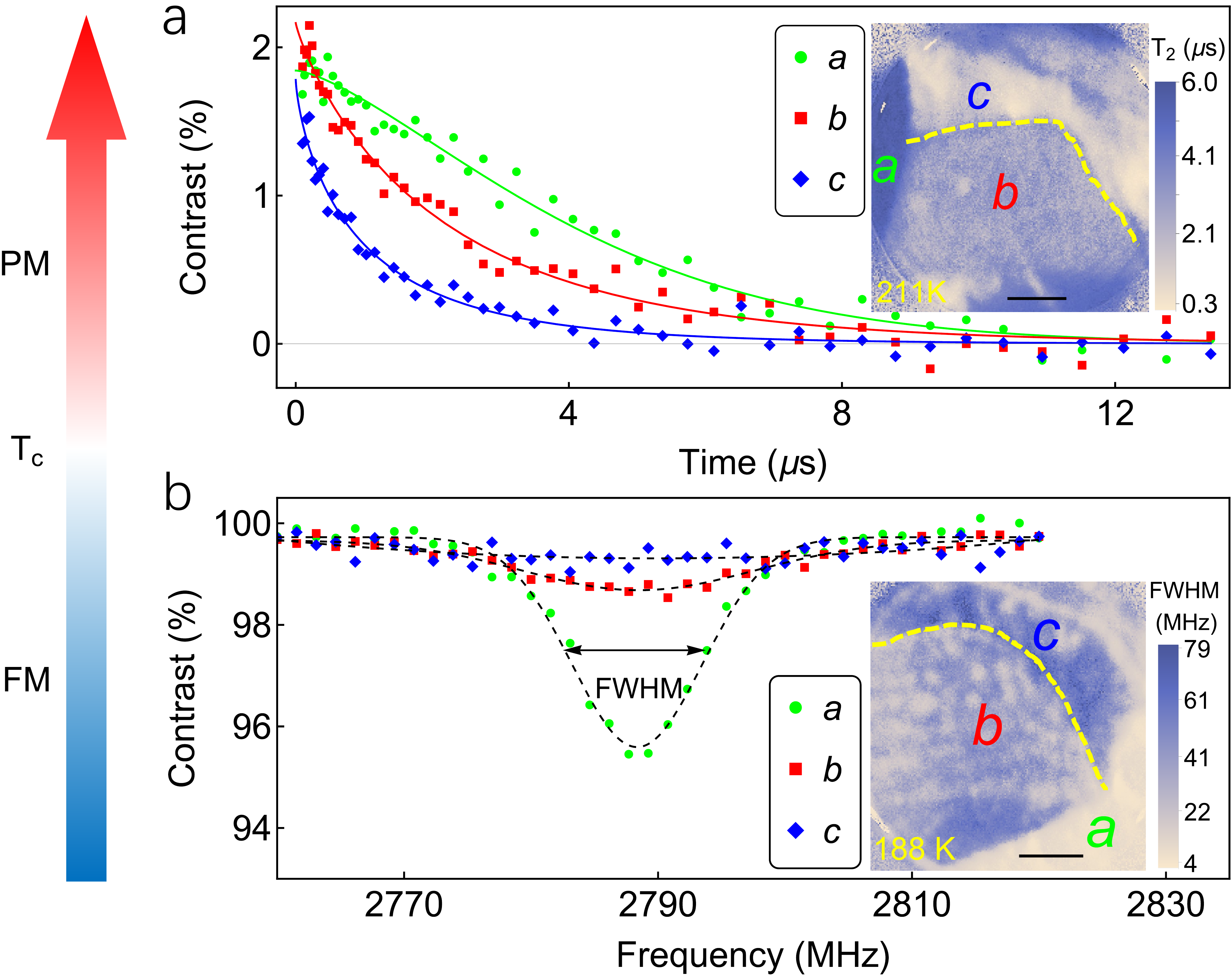}
		\caption{{\bf Dynamic and static magnetometry in the paramagnetic and ferromagnetic states}. (a) The typical decoherence curves of NV centers for the FGT sample in the paramagnetic (PM) phase with $T > T_c$. Solid lines are fitted curve using Eq. \ref{eq-CtC0}, yielding $\alpha = 1.306\pm 0.113$ (green), $0.772\pm 0.089$ (red), $0.698 \pm 0.098$ (blue) for the three curves. The inset shows the $T_2$ map at $T = 211$ K. See statistics of $\alpha$ and $T_2$ in Section \uppercase\expandafter{\romannumeral 6} of  Supplementary Material \cite{Supp_Mat}. (b) The CW spectrum of NV centers for the FGT sample in the ferromagnetic (FM) phase with $T < T_c$. Inset shows the FWHM map of the CW spectrum at 188 K ($T < T_c$). In both figures, the data are measured from  sample \#6, where region `a' without FGT sample, `b' and `c' are covered by FGT samples with thicknesses of 70 nm, and 90 nm, respectively. Representative curves in these three regimes are presented in the main panels. Scale bar: 5 $\mu$m.}
	\label{fig-fig2}	
 \end{figure}

Firstly, we perform wide-field imaging of the coherence time of the single-layer NV centers adjacent to the FGT sample. When $T > T_{c}$, the FGT is in paramagnetic state, exhibiting vanished magnetism. However, the local magnetic noise generated by the spin fluctuation can still influence the coherence of NV centers. The results are presented in Fig. \ref{fig-fig2} (a). Our sample contains three regions: `a' without sample, `b' with a 70 nm thick sample, and `c' with a 90 nm thick sample, which are determined by atomic force microscopy. The spatial distributions of $T_2$ in these regions are presented in the inset. By fitting the contrast using 
\begin{equation}
    C(t)=C_0 \exp(-\chi(t))= C_0 \exp(-(t/T_{2})^{\alpha}).
    \label{eq-CtC0}
   \end{equation}
where $C_0$ is the contrast derived from the difference between the signal frame and the reference frame and normalized to the reference frame. $T_{2}$ is the coherence time, which can be determined by $\chi(t) = 1$; and $\alpha$ is the stretch exponent related to the properties of noise. The three curves in Fig. \ref{fig-fig2} (a) show that the spin noise from the sample influences the corresponding $T_2$ and the exponent $\alpha$. In the region without sample, $\alpha \sim 1.5$, which is consistent with Ref. \cite{Bauch2020Decoherence}, while in the regions `b' and `c' covered by samples, $\alpha$ is greatly decreased. The stretch exponent $\alpha$ is a rather subtle issue in NV centers,
which can be affected by the inhomogenity of the sample \cite{Bauch2020Decoherence,Dobrovitski2008Decoherence,Park2022Decoherence,ghassemizadeh2024coherence}. 
In Fig. \ref{fig-fig2} (b) with $T<T_c$, the spontaneous magnetization happens and the CW spectrum (see Section \uppercase\expandafter{\romannumeral 4} of  Supplementary Material \cite{Supp_Mat}) is employed to image the static magnetic field generated by the sample. We find that a clear dip at about 2.78 GHz is observed in region `a' without sample below $T_c$. In contrast, in regions `b' and `c' the stray magnetic field from 
the FGT broadens the spectral lines and reduces the contrast, yielding FWHM of the spectra much larger than 12 MHz. In Section \uppercase\expandafter{\romannumeral 5} of  Supplementary Material \cite{Supp_Mat} for sample \#9, we show that the dips can be found when the domain walls are large enough. 

\begin{figure}[ht]
\centering
\includegraphics[width=1\linewidth]{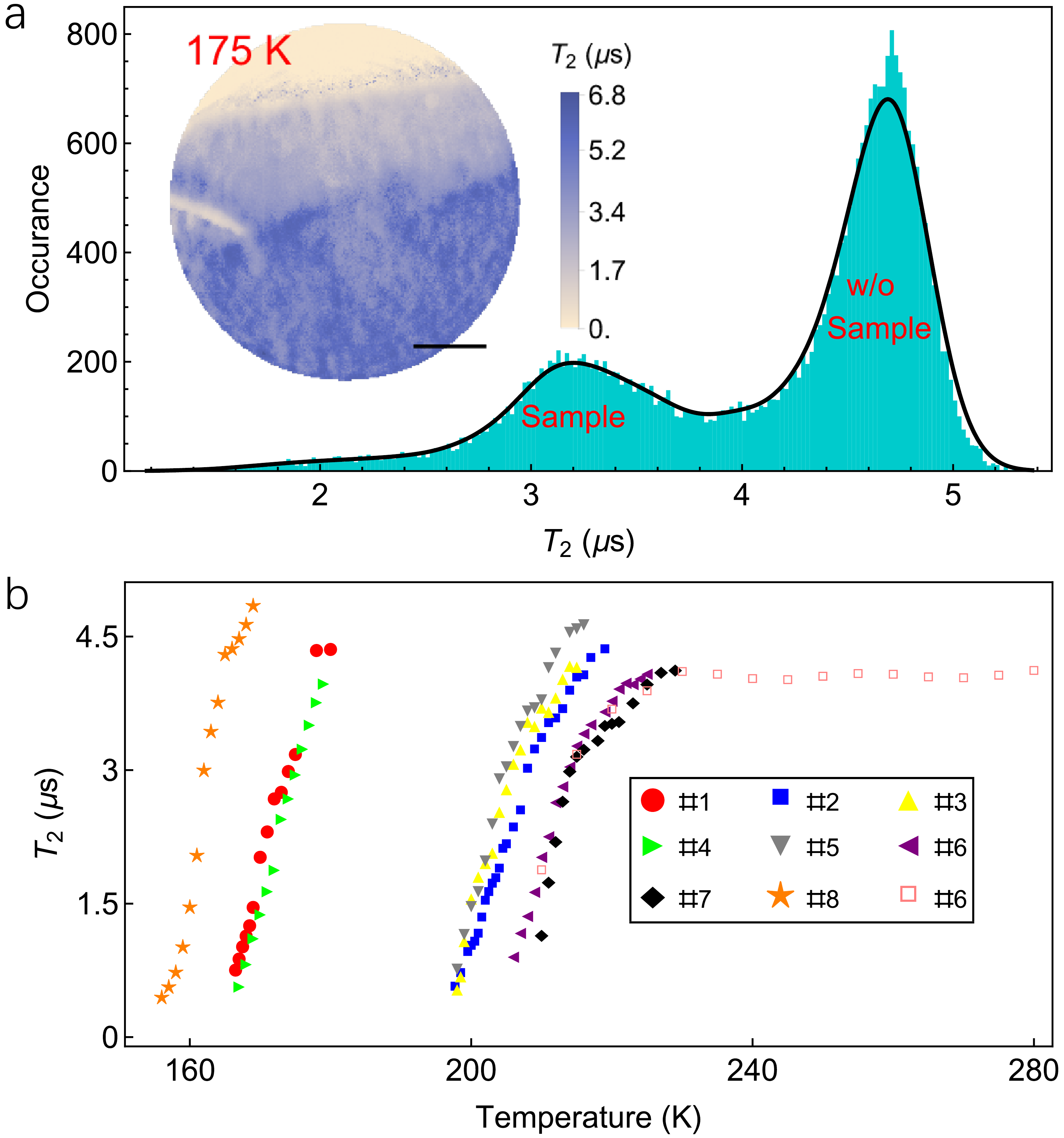}
\caption{{\bf Distribution of $T_2$ above $T_c$ and its temperature dependence}. (a) Histogram of the $T_{2}$ across the field of view. Inset shows the corresponding $T_{2}$ map measured from sample \#1 at $T = 175$ K. The histogram of $\alpha$ is presented in Section \uppercase\expandafter{\romannumeral 6} of  Supplementary Material \cite{Supp_Mat} .  (b) Coherence time $T_2$ for eight different samples \#1 - \#8 as a function of temperature. In our experiments, $T_{2}\le$ 500 ns can not be obtained because the number of pixels with satisfactory fit is insufficient for robust statistical analysis, thus is not shown. Scale bar: 5 $\mu$m.} 
\label{fig-fig3}  
\end{figure}

Next, we investigate the coherence time near and above the critical point. The distribution of $T_2$  is presented in Fig. \ref{fig-fig3} (a), showing two major peaks at  $\sim 3.2$ $\mu$s with the FGT sample and $ \sim 4.7$ $\mu$s without FGT sample. The distribution of $T_2$ comes from the inhomogeneity of magnetization in the sample. Furthermore, in Fig. \ref{fig-fig3} (b), we plot the peak of $T_2$ as a function of temperature for various samples with thicknesses of about 10 - 90 nm, showing that when $T$ approaches $T_c$ from above, all the measured $T_2$ gradually decrease to almost zero. The FGT samples with different thicknesses may have different $T_c$; however, all data exhibit the same feature. We have verified that the intrinsic coherence time of the NV centers is approximately 4.5 $\mu$s in our diamond chips \cite{Bauch2020Decoherence}, thus the dramatic decrease of coherence time should come from the magnetic noise induced by the critical fluctuation effect. Further details can be found in Section \uppercase\expandafter{\romannumeral 2} of  Supplementary Material \cite{Supp_Mat}.
\begin{figure}[ht] 
	\centering
	\includegraphics[width=1\linewidth]{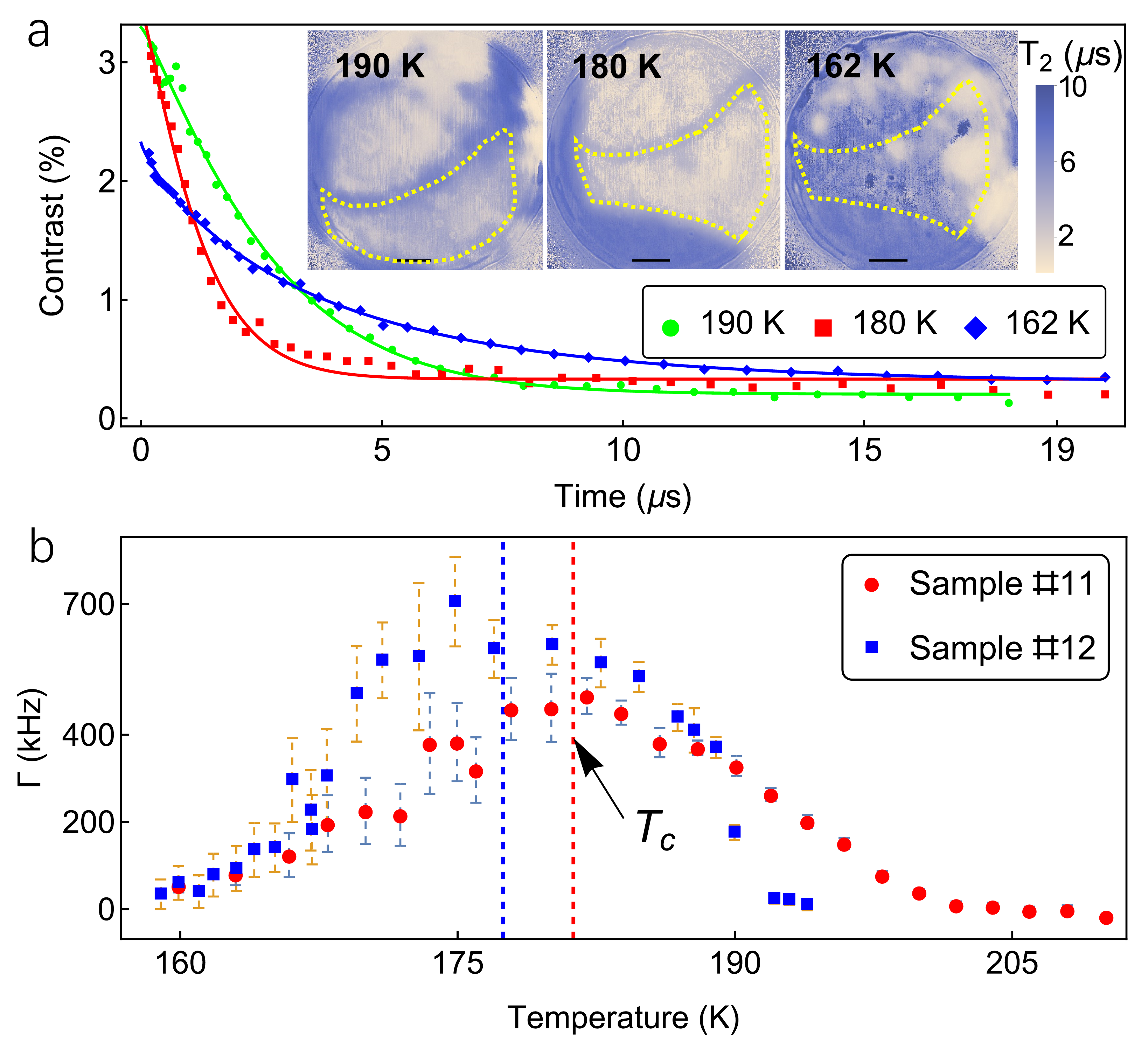}
	\caption{{\bf Coherence of NV centers across the critical point}. (a) The decoherence curve of the whole region with FGT sample (delineated by the green dashed line in Inset) at three different temperatures: above the $T_{c}$ (190 K, Green), near the $T_{c}$ (180 K, Red), below the $T_{c}$ (162 K, Blue). Solid lines are fitted curves using Eq. \ref{eq-CtC0}, yielding $\alpha = 1.183\pm 0.055$ (green), $1.187\pm 0.098$ (red), $0.845\pm 0.028$ (blue). Inset shows the $T_{2}$ map measured from sample \#12 at three temperatures.  (b) Decoherence rate $\Gamma$ of the NV centers across the critical temperature. The two samples have different thicknesses and different distances $d$ to NV centers: (47 nm, 310 nm for sample \#11) and (16 nm, 245 nm for sample \#12), respectively. Vertical dashed lines mark the $T_c$ estimated by the maximum. Scale bar: 5 $\mu$m.}
\label{fig-fig4}
\end{figure}

In the preceding two experiments, the distance $d$ from the FGT layer to the NV centers layer is about 60 nm \cite{ZIEGLER2010SRIM,Hauf2011NVSRIM} (without inserting the hBN flakes). When the $T_{c}$ is approached, the critical spin fluctuation is so strong that the coherence is completely destroyed. The statistical method presented in Fig. \ref{fig-fig3} (a) can not yield reliable results when $T_2$ is extremely short. We can mitigate this difficulty by inserting the hBN flakes with finite thickness between the diamond chip and the FGT sample, which increases the distance between the two layers thereby reducing the magnetic noise perceived by the NV centers. With this idea, the effect of thickness on the coherence rate can be performed in a single sample, with details presented in Section \uppercase\expandafter{\romannumeral 7} of  Supplementary Material \cite{Supp_Mat}. A rough estimation shows that the magnetic field strength as a function of distance $d$ scales as $\delta B \sim d^{-2.5}$. 

Therefore, by increasing the distance $d$ the measured spin noise will reduce near the critical point, enabling us to measure the $T_2$ across $T_c$. Fig. \ref{fig-fig4} (a) presents three decoherence curves at 190 K, 180 K, and 162 K, which corresponds to the situations above, near and below the $T_{c}$. Obviously, at about 180 K, the NV centers have the shortest $T_{2}$. It is worth noting that in the ferromagnetic state with $T < T_c$, there exists a pronounced magnetic noise variation within the FGT sample, as evident in the inset of Fig. \ref{fig-fig4} (a), then the data processing strategy in Fig. \ref{fig-fig3} (a) is no longer suitable. In this condition, we sum the data from all pixels within the region of interest to extract the characteristic coherence time. When $T > T_{c}$, these two data processing methods yield the same results, as shown in Section \uppercase\expandafter{\romannumeral 8} of  Supplementary Material \cite{Supp_Mat}. This also confirms the reliability of our experimental data, reflecting the overall characteristics of the sample rather than a local feature. The decoherence rate of NV centers for FGT sample \#11 and \#12 are presented in Fig. \ref{fig-fig4} (b) and the corresponding $T_{2}$ maps are presented in Fig.S8 and Fig. S9 in  Supplementary Material \cite{Supp_Mat}. Here the decoherence rate is defined as $\Gamma =1/T_2 - 1/T_{2,0}$, where $T_{2,0}$ is the intrinsic coherence time (without FGT sample). These two samples have different thicknesses and distances, yet they exhibit consistent results. In all measurements, we find that the decoherence rate exhibits a maximum value at the critical point, where the samples have the strongest critical fluctuations. This result is also consistent with the theoretical results in Ref. \cite{Chen2007Measurement}. 

Finally, with these observations, we are in a position to understand the fundamental physics involved. The two questions raised in the introduction can be properly addressed. The effective Hamiltonian of the NV centers should be written as \cite{Machado2023Quantum} 
\begin{equation}
H = \Delta_{0}(S_{z}^{2}-\frac{1}{3})+ \gamma (B_{0}+\delta B) S_{z},
\end{equation}
where $\Delta_{0} = 2.87$ GHz is the zero-field splitting for the NV centers, $B_{0}$ is the bias field and $\delta B$ is the fluctuation field generated by FGT substrate, which depends on the distance between the NV layer and FGT substrate (see Fig. \ref{fig-fig1}). The coefficient $\gamma =2.8$ MHz/Gs is the electron gyromagnetic ratio. This random field induces a pure dephasing process determined by $\langle \exp(i\phi(t))\rangle = \exp(-\frac{1}{2} \langle \phi(t)^{2}\rangle)$, where $\phi (t)=\gamma \int_{0}^{t} \delta B(t') d t'$. When the field $\delta B(t)$ is correlated, it yields some kind of $1/f$ noise, which is commonly presented in physical systems \cite{Machado2023Quantum, Claeys1997f,Liu2023f,bak1987f,Avinash2017f,RAYCHAUDHURI2002f}. 

Let $\langle \delta B(t_1) \delta B(t_2) \rangle = \frac{1}{2\pi} \int d\omega \exp(-i\omega(t_1 - t_2)) S(\omega)$, where $S(\omega)$ is the power spectra density of noise, then we have (see Eq. \ref{eq-CtC0}) 
\begin{equation} 
\label{eq:eq4}
\chi(t) \sim {1\over 2} \langle \phi^2(t)\rangle = {1 \over 4\pi} \int d\omega S(\omega) W_t(\omega),
\end{equation}
with frequency filter function for the Hahn-echo pulse sequence by $W_t(\omega) = 8\sin(\omega t/4)^4/\omega^2$ \cite{Degen2017QuantumSensing}. Thus for  noise with the $S(\omega) \sim 1/\omega^{\mu}$ ($\mu=0$ for white noise and $\mu \ne 0$ for $1/f$ noise), we have 
\begin{equation}\label{eq:eq5}
{1\over 2} \langle \phi^2\rangle \sim \int d\omega {\sin(\omega t)^4 \over \omega^{2}} {1\over \omega^{\mu}}\sim ({t \over T_2}) ^{1+\mu}.
\end{equation}

\begin{figure}[ht]
    \centering
    \includegraphics[width=1\linewidth]{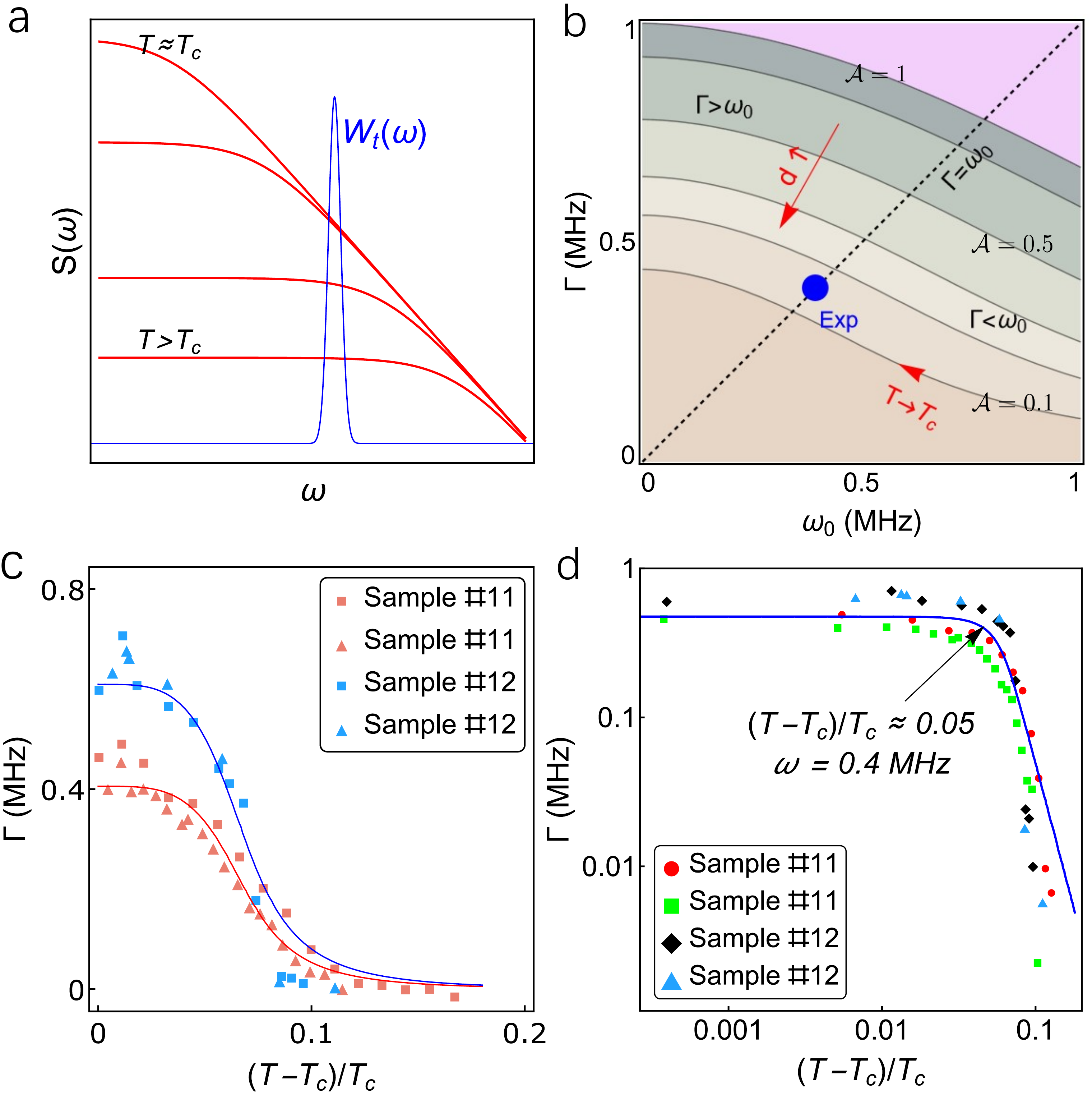}
 \caption{{\bf Theoretical understanding of the enhanced decoherence rate from noise spectra}. (a) The red solid line illustrates the spin fluctuation induced power spectral density $S(\omega)$ at different temperatures. The blue solid line represents the filter function of the Hahn-echo pulse sequence $W_t(\omega)$. (b) The contour plot of $A$ as a function of $\Gamma$ and $\omega_0$ for $\mathcal{A}$ = 0.1, 0.5 to 1 MHz$^{2.8}$.  The red arrows indicate the parameter change caused by increasing distance $d$ and approaching $T_{c}$, respectively. The blue point corresponds to the experiment result. The black dashed line marks the condition $\Gamma = \omega_{0}$, dividing the parameter space into two parts: $\Gamma > \omega_0$ and $\Gamma < \omega_0$. (c) Fitted $\Gamma$ (see Eq. \ref{eq:eq3}) as a function of $(T-T_c)/T_c$ for different samples. (d) The double-logarithmic plot of (c); and the black arrow is the crossover transition from white noise to the $1/f$ noise.}
    \label{fig-fig5}
\end{figure}

On the other hand, we anticipate the power spectral density function of critical magnetic fluctuations to exhibit the following form \cite{Chen2007Measurement}
\begin{equation} 
S(\omega) = A/(\omega_0^\mu + \omega^\mu),
\end{equation}
where the $\omega_{0} \sim |\frac{T-T_{c}}{T_{c}}|^{z\nu}$ is the reciprocal of the relaxation time, which is an intrinsic property of the sample, $z$ is the dynamic critical exponent associated with the relaxation time, and $\nu$ is the critical exponent associated with the divergence of the correlation length. For the 2D Ising model, $\mu=1.8$ \cite{Chen2007Measurement}. The $\omega_{0}$ is also the crossover frequency from white noise to $1/f$ noise, thus when $\omega \ll \omega_0$, it behaves like white noise, and when $\omega \gg \omega_0$, it gives $1/f$ noise, as depicted in Fig. \ref{fig-fig5} (a). Furthermore, the filter function $W_{t}(\omega)$ is peaked at $\omega t \simeq 1$, where $t$ is the interval of the pulse sequence. Thus only noise with $\omega \simeq 1/t$ will be retained, and the other frequencies will be filtered out. In this way, by varying the pulse intervals, we can measure the $S(\omega)$. According to Eq. \ref{eq:eq5}, for the white noise, we have $\frac{1}{2}\langle \phi^{2}\rangle \sim \frac{t}{T_{2}}$, thus in the fitting curve we have $\alpha \sim 1$. For the $1/f$ noise with $S(\omega) \sim 1/\omega^{\mu}$, we have $\frac{1}{2}\langle\phi^{2}\rangle \sim (\frac{t}{T_{2}})^{1+\mu}$, with $\alpha \sim 1+\mu$, thus $\alpha > 1$. Thus near the critical point, we should have an exponent different from $\alpha = 1$; see Fig. \ref{fig-fig2} using Eq. \ref{eq-CtC0}. Note that our measured exponent is different from the theoretical value of $\alpha = 2.8$ (with $\mu = 1.8$) due to the averaging effect \cite{Bauch2020Decoherence}; see statistics of $\alpha$ and $T_2$ in Section \uppercase\expandafter{\romannumeral 6} of  Supplementary Material \cite{Supp_Mat}. 

We are particularly interested in the decoherence rate $\Gamma$, which can be written as $\Gamma = 1/T_2$, ignoring of $T_{2,0}$ when $T_2 \ll T_{2,0}$. According Eq. \ref{eq:eq4}, we have $\chi(t)=(\Gamma t)^{\alpha} ={1 \over 4\pi} \int d\omega S(\omega) W_t(\omega)$, where 
$W_t(\omega)$ is a narrow function  compared with the spectral density $S(\omega)$, thus in this integral we can replace $t = 1/\Gamma$, yielding $1 = S(\Gamma) \int W_{1/\Gamma}(\omega) d\omega = S(\Gamma) {\pi \over 2\Gamma}$. This yields the following important linking between theory and experiment
\begin{equation}
\label{eq:eq3}
    S(\Gamma)=\frac{\mathcal{A}}{\omega_{0}^{\mu}+\Gamma^{\mu}}=\Gamma,
\end{equation}
where $\mathcal{A}$ is a constant depending on the distance $d$. This equation is one of the key results of this work. The relation between $\Gamma$, $\omega_0$ and distance is illustrated in Fig. \ref{fig-fig5} (b). For fixed $\mathcal{A}$ (thus fixed $d$), the limit $\omega_0 \rightarrow 0$ means approaching of $T_c$, which yields $1/f$ noise. With the increasing of $\omega_0$, $\Gamma$ should be reduced monotonically. Furthermore, increasing of $d$ will reduce $\mathcal{A}$ (since $\lim_{d\rightarrow \infty} \mathcal{A} = 0$), thus the increasing of $d$ will reduce $\Gamma$. Using the relation between $\omega_0$ and temperature $\omega_{0}=m|(T-T_{c})/T_{c}|^{z\nu}$, where $m$ is a fitting parameter, we can study the relation between $\Gamma$ and temperature. The results are presented in Fig. \ref{fig-fig5} (c), in which $\Gamma$ is from the experiment and $m$ and $\mathcal{A}$ are obtained by fitting the data using Eq. \ref{eq:eq3}; more discussions are presented in Section \uppercase\expandafter{\romannumeral 9} of  Supplementary Material \cite{Supp_Mat}. 

We find that this curve can fit all the data excellently. In this analysis, we have used $\nu = 1$ and $z = 2.17$ from the 
2D Ising model. Within this assumption, we have the following two intriguing limits 
\begin{equation}
\Gamma \sim
\left \{ 
\begin{aligned} 
& \text{constant} & \omega_0 \ll \Gamma \\ 
& |{T-T_c \over T_c}|^{-z\nu\mu}  & \omega_0 \gg \Gamma
\end{aligned}
\right.
\end{equation} 
These two limits and their crossover are presented in Fig. \ref{fig-fig5} (d), in which for all four measurements, they exhibit almost the same behavior. The crossover between these two noises is given by $\Gamma = \omega_0$. From these data we find at about $(T-T_c)/T_c = 0.05$, the relaxation time of the critical fluctuation is  about $1/\omega_0 \sim 2.5$ $\mu$s, which is independent of distance $d$. This point is also marked in Fig. \ref{fig-fig5} (b).

To conclude, we find that the $1/f$ noise near the  critical temperature and the white noise away from the critical point yield different dephasing process in strength to the NV centers, and the crossover for these two noises is realized by controlling the temperature and distance between the sample and the diamond. A theory for the relation between the critical fluctuation and the noise spectra is developed to understand the observations. In future, we hope that the critical temperature and the critical exponents  $\mu$, $z$, and $\nu$ determined or used in this way can be calibrated in more detail in conjunction with 
the other experimental methods, which can yield a much better understanding of criticality during phase transitions. 

{\it Acknowledgement}: This work is supported by the National Natural Science Foundation of China (grant No. T2125011), the CAS (grant No. GJJSTD20200001, YSBR-068, YSBR-049), Innovation Program for Quantum Science and Technology (Grant No. 2021ZD0302200, 2021ZD0303204, 2021ZD0301200, 2021ZD0302800 and 2021ZD0301500),  the National Key Research and Development Program of China (Grant No. 2023YFB4502500), New Cornerstone Science Foundation through the XPLORER PRIZE, and the Fundamental Research Funds for the Central Universities.

\bibliography{Pap_liyx.bib}
\end{document}